\documentclass[conference,letterpaper]{IEEEtran}

\usepackage{graphicx}
\usepackage{algorithm, ,algorithmic}
\usepackage{threeparttable}
\usepackage[utf8]{inputenc} 
\usepackage[T1]{fontenc}
\usepackage{url}
\usepackage{ifthen}
\usepackage{cite}
\usepackage{diagbox}
\usepackage{caption3}
\usepackage[cmex10]{amsmath} 

\begin{document}
\title{{Thermodynamically Stable DNA Code Design using a Similarity Significance Model}}

\author{%
   \IEEEauthorblockN{Yixin Wang\IEEEauthorrefmark{1},
                     Md Noor-A-Rahim\IEEEauthorrefmark{2}, Erry Gunawan\IEEEauthorrefmark{1}, Yong Liang Guan\IEEEauthorrefmark{1},
                     and Chueh Loo Poh \IEEEauthorrefmark{3}\IEEEauthorrefmark{4}}
   \IEEEauthorblockA{\IEEEauthorrefmark{1}%
                     School of Electrical \& Electronic Engineering, Nanyang Technological University, 639798, Singapore,
                     \\\{ywang065, egunawan, eylguan\}@ntu.edu.sg}
   \IEEEauthorblockA{\IEEEauthorrefmark{2}%
                     School of Computer Science \& IT, University College Cork, Ireland,
                     m.rahim@cs.ucc.ie}
   \IEEEauthorblockA{\IEEEauthorrefmark{3}%
                     Department of Biomedical Engineering, National University of Singapore (NUS), 117583, Singapore,
                     }
   \IEEEauthorblockA{\IEEEauthorrefmark{4}%
                     Synthetic Biology for Clinical and Technological Innovation, NUS, Singapore.\\poh.chuehloo@nus.edu.sg}
 }

\maketitle



\begin{abstract}
DNA code design aims to generate a set of DNA sequences (codewords) with minimum likelihood of undesired hybridizations among sequences and their reverse-complement (RC) pairs (cross-hybridization).  Inspired by the distinct hybridization affinities (or stabilities) of perfect double helix constructed by individual single-stranded DNA (ssDNA) and its RC pair, we propose a novel similarity significance (SS) model to measure the similarity between DNA sequences. Particularly, instead of directly measuring the similarity of two sequences by any metric/approach, the proposed SS works in a way to evaluate how more likely will the undesirable hybridizations occur over the desirable hybridizations in the presence of the two measured sequences and their RC pairs. With this SS model, we construct thermodynamically stable DNA codes subject to several combinatorial constraints using a sorting-based algorithm. The proposed scheme results in DNA codes with larger code sizes and wider free energy gaps (hence better cross-hybridization performance) compared to the existing methods.
\end{abstract}


\section{Introduction}
DNA code is an ensemble of $q$-ary ($q=4$) $n$-sequences subject to combinatorial biological constraints and with controlled maximum similarity among sequences and their reverse-complement (RC) pairs. Such sequences are sought after for wide applications including DNA computing \cite{adleman1994molecular}, DNA memory, and DNA data storage \cite{organick2018random, wang2019high}. The theoretical bound of DNA codes satisfying combinatorial biochemical constraints, such as GC content, word-word distance, and word-reverse complementary word distance have been explored in \cite{marathe2001combinatorial, king2003bounds, chee2008improved}. Concurrently, several code construction approaches have been proposed, including template-based construction \cite{arita2002dna, kobayashi2002template, king2007binary}, search algorithm-based construction \cite{deaton1996genetic,tulpan2002stochastic, wang2018constructing}, and coding theoretic construction \cite{gaborit2005linear, milenkovic2005design}.
 
 

In DNA code design, one major criterion to be controlled is the maximum similarity or minimum distance between DNA sequences or/and their reverse-complement (RC) sequences (i.e., DNA strands in the physical entity). This criterion has a significant impact on the cross-hybridization performance of a DNA code that can be evaluated thermodynamically. Conventionally, thermodynamic-based models and distance models have been used for DNA code construction, of which the latter approach is recognized with lower complexity \cite{phan2009codeword}. In particular, Hamming distance model has been widely used in theoretic-based code design \cite{gaborit2005linear,marathe2001combinatorial, king2003bounds, chee2008improved, tulpan2002stochastic, kawashimo2006dna}, while edit distance model has been either adopted in a straight-forward fashion \cite{wang2018constructing, d2005new, bennenni2019greedy} or generalized to pairwise alignment algorithms \cite{altschul1990basic} that are widely implemented in biological software \cite{wang2003selection, nielsen2003design, xu2009design, ye2012primer}. In a nutshell, the existing similarity/distance models might discriminate in obtaining or quantifying the similarity value while they are all built upon directly comparing two unique sequences that are under consideration. In contrast, we introduce a new similarity significance (SS) model that involves not only the two compared sequences but also their RC pairs. In this model, the RC pairs are considered as the reference pairs for quantifying the affinity-mediate similarity of the two comparing sequences.

With the proposed SS model, we designed DNA codes using a maximum SS constraint to restrain the cross-hybridization performance of a set of synthesized DNA strands (i.e., the code). Specifically, by using SS, the similarity of two different DNA strands (e.g., $u, v$) is determined by evaluating how more likely will one strand ($u$ or $v$) hybridize with the RC pair of the other strand (denoted by $v'$ or $u'$) than hybridizing with its RC pair ($u'$ or $v'$). This differs from the classic sequence comparison approach where distance/similarity metrics are directly used to determine how similar two DNA sequences are. To quantify the SS of two sequences, we used a sequence alignment between two sequences with a biased similarity weight allocation for DNA symbols that are identical under the alignment. By a sorting-based exhaustive search algorithm, the DNA codes are attained. The constructed DNA codewords satisfy the predefined maximum SS parameters while complying with the balanced GC content and maximum homopolymer run. The existence of codes with larger code sizes and wider free energy gaps (better cross-hybridization performance) using the SS model rather than traditional metrics and existing measure implies a better approximation towards the thermodynamic property.

\section{Definition, Notation}
A hybridization conformation is a double helix formation which is produced from single-stranded DNA (ssDNA) strand(s) with two chemically distinct ends (i.e., 5'- end and 3'- end) under the Watson-Crick (WC) complement rule where adenine (A) binds with thymine (T) and cytosine (C) binds with guanine (G). For a DNA sequence denoted by $u$, the desirable hybridization is the perfect hybridization that occurs between $u$ and its WC RC pair (denoted by $u'$). For instance, $u$ = 5'-TTCCGAT-3' and $u'$ = 5'-ATCGGAA-3'. All other $u$-involved hybridizations are considered undesirable to $u$, rendering the cross-hybridization. The stability of a hybridization conformation could be predicted by a thermodynamic metric, termed Gibbs free energy, in which a lower value (usually negative) implies a more stable/possible helix conformation \cite{phan2009codeword}. As such, the hybridization affinity of DNA strand(s) could be inferred by the minimum free/Gibbs energy (MFE) that stands for the most possible conformation structure of the considered ssDNA strand(s). For simplicity, $\Delta G$ is used to denote MFE in the following context. 

Consider a DNA code $\mathcal{C}$ consisting of unique sequences (e.g., $u, v \in \mathcal{C}$), there are three categories of undesirable hybridizations that might occur to sequence $u$. First, a self-complement hybridization that occurs within $u$, e.g., $u$ folds with itself. Second, a sequence-sequence (tag-tag) hybridization, which occurs between $u$ and the other candidate ssDNA strand $v$. Third, a sequence-RC (tag-target) hybridization, in which $u$ hybridizes with the RC pair of the other candidate ssDNA strand ($v'$).

MFE is widely used to characterize the hybridization affinity between two ssDNA strands. To estimate the cross-hybridization performance of a DNA code (i.e., a set of DNA sequences), generalized thermodynamic criteria based on MFE have been proposed in \cite{shortreed2005thermodynamic, zhang2013novel}. Similarly, we use the \textit{free energy gap} ($\delta$) to evaluate a DNA code, which is defined by,

\begin{eqnarray}
\label{eq:free energy gap}
\setlength\belowdisplayskip{-10pt}
\begin{aligned}
    &&\delta = \min\limits_{u\in\mathcal{C}}\{\min\limits_{v\in\mathcal{C}, v \neq u}\{\Delta G(u,u),\Delta G(u,v),\\
    &&\Delta G(u,v'),\Delta G(u',v')\}-\Delta G(u,u')\}
\end{aligned}
\end{eqnarray}

\noindent{where $\Delta G(u, v)$ represents the MFE of two unique sequences $u$ and $v$ that pertain to a DNA code $\mathcal{C}$, and $u'$ and $v'$ are the WC RC pairs of $u$ and $v$, respectively. In \eqref{eq:free energy gap}, the first three $\Delta G$s within the inner $\min\{.\}$ correspondingly refer to the MFE of three undesirable hybridizations of strand $u$ as mentioned, i.e., self-fold hybridization ($\Delta G(u, u)$), sequence-sequence hybridization ($\Delta G(u, v)$), and sequence-RC hybridization ($\Delta G(u, v')$). Conversely, the $\Delta G(u, u')$ out of the inner $\min\{.\}$ represents the MFE of the desirable hybridization of strand $u$ (i.e., the complete duplex formation with its RC pair $u'$). Note that $\Delta G(u', v')$ which indicates the MFE of the undesirable hybridization constructed by the RC pairs of codeword strands (i.e., $u'$ and $v'$) is also included in (\ref{eq:free energy gap}) to keep consistency with \cite{zhang2013novel}. Notably, $\delta$ used in this work is more stringent than the free energy gaps used in \cite{zhang2013novel} and \cite{shortreed2005thermodynamic}, as their definitions exclude the MFE of self-fold hybridizations  (i.e.,$\Delta G(u, u)$) and RC-RC hybridizations (i.e., $\Delta G(u', v')$), respectively. In this work, $\delta$ is calculated based on the MFEs attained from the online tool DINAMelt \cite{markham2005dinamelt}. According to \eqref{eq:free energy gap}, a larger $\delta$ represents a wider gap between the free energies of the desirable and undesirable hybridizations, and thus indicating a better DNA code. Therefore, the free energy gap $\delta$ defined in \eqref{eq:free energy gap} can be used to evaluate the cross-hybridization performance of a DNA code.}

\section{Proposed similarity significance (SS) model}
The proposed similarity significance (SS) is inspired by the distinct affinities of individual perfect hybridization (between the WC complement pair, i.e., an ssDNA and its RC) and the definition of the free energy gap of a DNA code (i.e., the MFE difference between undesirable hybridizations and desirable hybridizations). 

Unlike the conventional measures (e.g., Hamming distance) which quantify the distance/similarity of two different DNA sequences, the proposed SS quantifies the significance of the \textit{mutual similarity} between two different sequences (which is undesirable) over the \textit{self similarity} between identical sequences or to say sequences with themselves (which is desirable). This model inherently aligns with the thermodynamic metric of the code, i.e., the free energy gap between the undesirable hybridizations and desirable hybridizations. Specifically, $\tau(u,v)$ is attained by computing the ratio of the similarity of sequences $u$ and $v$ against the similarity of the sequences $u$ and $u$ or the similarity of sequences $v$ and $v$. Naturally, $\tau(u,u)=\tau(v,v)=1$, which shares the same essence with the normalization in this respect. However, unlike the normalized similarity where the normalization coefficient, i.e., the denominator, is usually fixed as the length of the sequences or resolved by the sequence alignment, the denominator in SS depends on the individually varied affinities of the perfect WC conformations that relate to the compared sequences.

By definition, we compare the proposed SS model with the traditional similarity models. Consider two sequences $s_1, s_2$, the similarity measured by a standard similarity model can be formed by
\begin{eqnarray}
    \tau_{\text{std}}(s_1,s_2) = \xi(s_1, s_2)
\label{eq:standard model}
\end{eqnarray}
\noindent{in which $\xi(s_1, s_2)$ is obtained by comparing sequences $s_1$ and $s_2$ with any customized similarity model, such as Hamming model and Edit model.}

Likely, the similarity measured by a normalized similarity model can be formed by
\begin{eqnarray}
    \tau_{\text{norm}}(s_1,s_2) = \frac{\xi(s_1, s_2)}{M(s_1, s_2)}
\label{eq:normalized model}
\end{eqnarray}
\noindent{in which $\xi(s_1, s_2)$ is same as (\ref{eq:standard model}) and $M(s_1, s_2)$ is the normalization parameter that might depend on the alignment or the length of the two comparing sequences.}

In contrast, the similarity measured by the SS model is formed by
\begin{eqnarray}
    \tau(s_1,s_2) = \frac{\xi_{\text{w}}(s_1, s_2)}{\text{max}\{\xi_{\text{w}}(s_1, s_1), \xi_{\text{w}}(s_2, s_2)\}}
\label{eq:SS model}
\end{eqnarray}
\noindent{in which $\xi_{\text{w}}(.,.)$ is obtained by comparing two sequences similar as $\xi(.,.)$ in (\ref{eq:standard model}) while it differs from (\ref{eq:standard model}) in quantification. To obtain $\xi_{\text{w}}(.,.)$ in (\ref{eq:SS model}), a weighted similarity allocation (WSA) mechanism is indispensable. The WSA is built upon the biochemical characteristics, and it is of critical importance as it discriminates the SS from the normalized similarity.}

\subsection{Quantifying similarity using SS model}
The key concept of SS is to measure the similarity of two unique DNA sequences by quantifying the relative likelihood of undesirable hybridizations against perfect WC hybridizations with the presence of both sequences and their RC pairs. Particularly, the likelihood of hybridizations could be measured by thermodynamic models or approximated by any similarity model that reflects the distinct hybridization affinities. Hence, any existing measure that could differentiate the affinity of hybridizations among sequences could be used as the basis of SS computation \cite{altschul1990basic, d2004weighted, bishop2007free, macula2008new, d2014dna}. Here, we use a best alignment criteria (BAC) and a weighted similarity allocation (WSA) for quantifying similarity using SS, where the BAC initials the base-pair comparison, and the WSA discriminates the SS from the normalized similarity. There are four steps in SS computation. First, a BAC will be applied to determine the best alignment between two compared sequences. Then, based on the WSA mechanism, the similarity score is attained under the best alignment. Next, the similarity scores of each sequence with itself are calculated using the WSA mechanism. Lastly, the ratio of the score in step 2 to the minimum score in step 3 is calculated as the result. Before presenting the formal formulation of the proposed SS, the BAC and the WSA that form the basis of SS computation are briefly explained as follows.


The \textit{best alignment criterion (BAC)} determines the best alignment of two DNA sequences (e.g., $u$ and $v$) in terms of approximating the alignment with which the most potential hybridization (between $u$ and $v'$) occurs. In an alignment, $l$ and $k$ denote the length of consecutively identical bases and the number of other identical bases between two compared sequences, respectively. With the insight of the biased effect of the continuous similarity $l$ over the discontinuous similarity $k$ on the conventional sequence alignment methods (e.g., BLAST \cite{altschul1990basic}), a BAC with priority to $l$ is used. Specifically, we derive the BAC by maximizing the continuous similarity while minimizing the dissimilarity using the objective function $f=l-(n-l-k)$, where $n$ is the length of the sequences, and $n-l-k$ is the dissimilarity. Let $L, K$ be the corresponding values of $l, k$ under the best alignment, the criterion can be summarized as,
%
\begin{eqnarray}
(L,K)=\underset{(l,k)}{\arg\max}f(l,k)=\{(l,k) \mid f(l,k) = 2l+k \}
\label{eq:newbestalignment}
\end{eqnarray}

The \textit{weighted similarity allocation (WSA)} mechanism gives biased similarity scores to aligned base pairs that compose bases from the compared sequences. The aligned base-pair can be identified as identical base-pair or distinct base-pair. Only identical base-pairs are given non-zero similarity weights. Moreover, considering that the binding between G and C is tighter than A and T due to the three hydrogen bonds existing between G and C, the WSA allocates additional similarity weights to identical G or C base-pair whose last (or left) aligned base-pair is also an identical base-pair. Specifically, two additional weights $\alpha$ and $\beta$ are assigned to identical G/C base-pair whose last aligned base-pair is identical G/C base-pair and identical A/T base-pair, respectively. Henceforward, $\alpha$ and $\beta$ are termed as \textit{consecutive G/C weight} and \textit{non-consecutive G/C weight}. Owing to this, the BAC related $l,k$ in (\ref{eq:newbestalignment}) becomes $l(\alpha, \beta), k(\alpha, \beta)$, where $\alpha\in[0,1]$ and $\beta\in[0,1]$ are as defined. Hence, the similarity vector $\vec{S}(u,v)$ of each compared sequence pair ($u,v$) can be derived under the best alignment that has been determined by the BAC. 

\begin{figure}[htbp]
   \centering
    \includegraphics[width=0.45\textwidth]{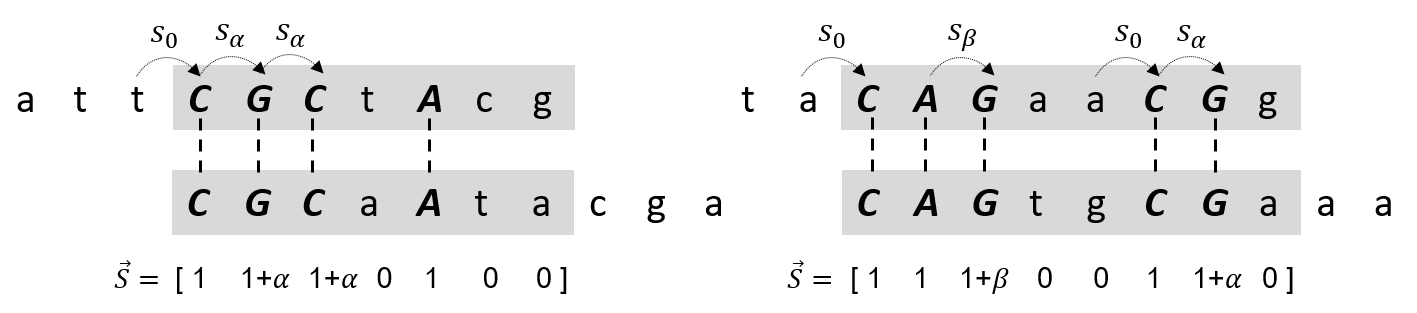}
   \caption{Two examples of similarity vectors under best alignments.}
   \label{fig:similarity vector}
 \end{figure}

The entries of the vector $\vec{S}(.,.)$ are the similarity weights of the aligned base-pairs from left to right of the alignment (excluding the two overhang ends). Note that values of the additional weights $\alpha,\beta$ in the (undesirable) similarity vector of unique sequences (i.e., $\vec{S}(u,v,\alpha_1,\beta_1)$) and the (desirable) similarity vector of identical sequences (i.e., $\vec{S}(u,u,\alpha_2,\beta_2)$), can be inconsistent (i.e., $\alpha_1 \neq\alpha_2;\beta_1 \neq\beta_2$). Two instances of the similarity vectors are shown in Figure~\ref{fig:similarity vector}, in which the best alignment region is gray-shadowed and the identical base-pairs under the best alignment are bold and italic. The $s_0$, $s_\alpha$, and $s_\beta$ correspondingly denote the state information transferred from the last aligned base-pair to the identical G/C base-pair for regulating the weight allocation. Specifically, $s_0$ indicates that there is no aligned base-pair in the left or the last/left aligned base-pair is not identical, thus no extra weight is added to the comparing identical G/C base-pair; $s_\alpha$ indicates that the left aligned base-pair is identical G base-pair or C base-pair, thus the \textit{consecutive G/C weight} $\alpha$ is added to the comparing identical G/C base-pair; while $s_\beta$ indicates that the left is an identical A or T base-pair, thus the \textit{non-consecutive G/C weight} $\beta$ is added to the comparing base-pair. 

With the similarity vector under the best alignment of two comparing sequences, the similarity score is obtained by accumulating the entries of the vector. According to the above-discussed definition of the SS, the SS $\tau(u,v)$ of two different sequences (i.e., $u\neq v$) of length $n$ is the ratio of undesirable similarity score (between $u$ and $v$ ) against the minimal desirable similarity score (between $u$ and $u$ or $v$ and $v$), and can be formulated by,
\vspace{-2pt}
\begin{eqnarray}
\begin{aligned}
    &\tau(u,v)=\tau(v,u) \\ & =\frac{\sum\limits^{n-\epsilon}\vec{S}(u,v,\alpha_1,\beta_1)}{\min \{\sum\limits^{n} \vec{S}(u,u,\alpha_2,\beta_2), \sum\limits^{n}\vec{S}(v,v,\alpha_2,\beta_2)\}}
\end{aligned}
\label{eq:similarity significance}
\end{eqnarray}

\noindent{where $\sum\vec{S}(\cdot)$ denotes the accumulation of values in vector $\vec{S}(\cdot)$, in which the size of the vector is $n-\epsilon$, where $\epsilon$ is the  length of overhanging bases with the best alignment. Note that for $\vec{S}(u,u,\alpha_2,\beta_2)$ and $\vec{S}(v,v,\alpha_2,\beta_2)$, $\epsilon=0$. }


\section{DNA codes with combinatorial constraints}
We design DNA codes having combinatorial constraints using a sorting-based exhaustive search.

\subsection{Code with balanced GC content and maximum homopolymer run constraints}
DNA sequences composing a balanced number of 'G'/'C' and 'A'/'T' symbols are desired by the parallel biological reactions because they are like to have a unified melting temperature ($T_m$). Besides, the maximum homopolymer run constraint that restricts the maximum allowable length of the consecutively repetitive symbols in the DNA sequence (named as continuity in \cite{shin2002evolutionary,shin2005multiobjective, zhang2013novel}), is also desirable for DNA code. Our previous result in \cite{wang2019construction} has shown an efficient construction of codes subject to these two constraints.



\subsection{Code with minimum distance/maximum similarity}
The minimum distance (or maximum similarity) between sequences has been used as a constraint for the DNA code design \cite{gaborit2005linear, marathe2001combinatorial, king2003bounds, chee2008improved, tulpan2002stochastic, kawashimo2006dna, d2005new}. The sequence-sequence distance (SSD) and sequence-RC distance (SRCD), which correspondingly imply the sequence-RC hybridization affinity and sequence-sequence/RC-RC/self-fold hybridization affinity, are required to satisfy the predefined constraint. In this work, a similarity significance (SS) ($\tau$) is introduced to measure the similarity of two DNA sequences. Therefore, similar to previous works \cite{kawashimo2006dna,phan2009codeword}, the SS ($\tau$) could be categorized into the \textit{sequence-sequence SS (SSSS)} and \textit{sequence-RC SS (SRCSS)}. As such, the DNA code design problem can be summarized as searching for a set of constrained DNA sequences with length $n$ ($\mathcal{C} \subseteq A_4^n$) such that,
\begin{eqnarray}
\setlength{\belowdisplayskip}{-20pt}
\left\{
\begin{array}{rcl}
    \max\limits_{u, v \in \mathcal{C}, u\neq v}\tau(u, v)\leq T_{th} &  \\ 
    \max\limits_{u, v \in \mathcal{C}}\tau(u, v')\leq T_{th} &
\end{array}  
\right.
\label{eq:similarity}
\end{eqnarray}

\noindent{where $A_4^n$ is the complete set of sequences of length $n$ with symbols from an alphabet set \{A, T, C, G\}; $\tau(u,v)$  and $\tau(u,v')$ are SSSS and SRCSS, respectively, with $u, v \in \mathcal{C}$; $T_{th}\in[0,1]$ is a predefined maximum SS parameter.}

With the approximated mapping existing in the SS measure $\tau$ and the MFE $\Delta G$, i.e.,$\tau(u,u')\leftrightarrow \Delta G(u,u); \tau(u,v')\leftrightarrow \Delta G(u,v),\Delta G(u',v'); \tau(u,v)\leftrightarrow \Delta G(u,v')$, the free energy gap $\delta$ defined in (\ref{eq:free energy gap}) could be approximated using a SS gap $T_{gap}$ with a renewed formula in terms of $\tau$ as follows,

\begin{eqnarray}
\begin{aligned}
    & T_{gap} = \max\limits_{u\in\mathcal{C}}\{\max\limits_{v\in\mathcal{C}, v \neq u}\{\tau(u,u'),\tau(u,v'),\tau(u,v)\} \\ &-\tau(u,u)\}.
\end{aligned}
\label{eq:similarity gap}
\end{eqnarray}

\noindent{Since $\tau(u,u)=1$, (\ref{eq:similarity gap}) becomes $T_{gap}=T_{th}-1$ for a DNA code generated based on (\ref{eq:similarity}). Therefore, either $T_{th}$ or $T_{gap}$ can be set in advance for designing a DNA code. For simplicity, maximum SS $T_{th}$ was used as the design parameter in our DNA code design. }


\subsection{An exhaustive search of constrained DNA codes}
The initial candidate constrained DNA sequences are generated using our method in \cite{wang2019construction}. This constrained sequence initialization outperforms the random sequence initialization of most evolutionary search algorithms \cite{cervantes2013improving,shin2005multiobjective} in terms of avoiding much search complexity. Leveraging by the initialization, the DNA code design problem is turned into an exhaustive search problem with the maximum SS $T_{th}$ as the only constraint. 

The following notations are used in Algorithm \ref{Algo: Search}:
\begin{itemize}
\item $\mathcal{S}(n, w_{gc}, k)$: A set of $n$-length DNA sequences subject to GC content $w_{gc}\in (40\%, 60\%)$ and maximum homopolymer run $k=3$.

\item $\mathcal{C}(n,T_{th})$: A set of n-length $(w_{gc},k)$ constrained DNA sequences with maximum SS $T_{th}$, i.e., DNA codes. 

\item $|.|$: The cardinality of a set.

\item $s_i...s_j$: The sub-string starting from $i$th symbol to $j$th symbol of sequence $S$.

\item $\mathcal{S}_x$: A subset of $\mathcal{S}(n, w_{gc}, k)$ built by grouping sequences with the same suffix, where x is an index indicator.

\item $\mathcal{S}_x \leftrightarrow \mathcal{S}_y$: A neighboring group pair built by linking the groups with the minimum Hamming distance in the suffix. 

\end{itemize}

 \begin{algorithm}
 \caption{An exhaustive search of DNA codes}
 \label{Algo: Search}
 \begin{algorithmic}[1]
 \renewcommand{\algorithmicrequire}{\textbf{Input:}}
 \renewcommand{\algorithmicensure}{\textbf{Output:}}
 \REQUIRE $\mathcal{S}(n, w_{gc}, k)$, $T_{th}$
 \ENSURE $\mathcal{C}(n,T_{th})$
 \\ \textit{Initialisation}: $\mathcal{S}=\mathcal{S}(n, w_{gc}, k), \mathcal{C}=\{\}$
 \\ \textit{Pre-processing:}
 \FORALL{$S \in \mathcal{S}$ such that $s_{n-\lfloor n\cdot T_{th}\rfloor}...s_n$ are equal}
\STATE Grouping $S$ into $\mathcal{S}_x$
\ENDFOR
\STATE Sort the groups into $\sum\mathcal{S}=\{\mathcal{S}_{1}, \mathcal{S}_{2}, ...\}$ with an order of increasing $|\mathcal{S}_x|$
\STATE Link each $\mathcal{S}_x \in \sum\mathcal{S}$ a neighboring group pair $\mathcal{S}_y$
 \\ \textit{Searching:}
\FOR{$x=1$ to $|\sum\mathcal{S}|$}
\STATE Select a sequence $X \in \mathcal{S}_x$.
\IF {$X$ satisfies (\ref{eq:similarity})}
\STATE Add the valid $X$ to $\mathcal{C}$, and eliminate $\mathcal{S}_x$ from $\sum\mathcal{S}$
\STATE Select a sequence $Y \in \mathcal{S}_y$, and verify the validity
\STATE Update $\mathcal{C}, \sum\mathcal{S}$ as line 9 if valid
\ELSE
\STATE back to line 7
\ENDIF
\ENDFOR
\STATE $\mathcal{C}(n,T_{th})=\mathcal{C}$
 \end{algorithmic} 
 \end{algorithm}

\section{Result}

 Given a specific sequence length, we investigate the code size and cross-hybridization performance of the codes constructed using different similarity/distance models. All codes are generated following the same searching process as discussed above except using different models as the similarity/distance measures of two sequences. The cross-hybridization is reflected by the free energy gap of $\delta$. The MFEs related to the $\delta$ were obtained from DINAMelt by setting the hybridization temperature to the universal $37^\circ$C. Note that as lower MFE values (negative) imply higher hybridization potentials, for convenience, the MFEs with positive values indicating nearly no hybridization potential are set to 0.

\begin{table}[!h]
   \setlength{\tabcolsep}{1.8mm}
   \centering
   \caption{Comparing free energy gap (code size) with Hamming model}\label{Tab:distance effect}
    \begin{tabular}{|l|l|l|l|l|l|}
    \hline
    \diagbox{Model}{$n$}  & 6 & 7 & 8 & 9 & 10 \\\hline
    Hamming & 1.0 (4)  & 2.1 (9) &  2.1 (22) & 2.3 (18) & 3.7 (15) \\ \hline
    SS & 2.4 (5) & 2.1 (11) & 2.3 (32) & 3.5 (18) & 4.8 (15)\\ \hline
    \end{tabular}
 \end{table}

\subsubsection{Comparison with Hamming-based codes}
  From Table~\ref{Tab:distance effect} where free energy gaps of the codes and the code sizes (parenthetical data) are shown, it could be found that, for all given lengths, the SS model enables codes constructed with no less free energy gaps and code sizes simultaneously. This indicates that by using the SS model, more DNA sequences with a given length could be generated with no adverse effect on the cross-hybridization performance. The thresholds used for different models might be varied provided that the codes generated are comparable in terms of code size and energy gap. The overall thresholds are chosen to avoid that the code sizes are too large to calculate the energy gaps. Notice that the energy gaps are currently calculated manually based on the free energy obtained from DINAMelt. 

\begin{table}[!h]
   \setlength{\tabcolsep}{1.8mm}
   \centering
   \caption{Comparing free energy gap (code size) with Edit model}\label{Tab:distance effect 2}
    \begin{tabular}{|l|l|l|l|l|l|}
    \hline
    \diagbox{Model}{$n$}  & 6 & 7 & 8 & 9 & 10 \\\hline
    Edit & 2.7 (3)  & 1.1 (6) &  2.4 (13) & 4.2 (10) & 5.4 (7) \\ \hline
    SS & 3.0 (3) & 2.4 (6) & 2.7 (13) & 4.7 (10) & 6.4 (7)\\ \hline
\end{tabular}
 \end{table}

\subsubsection{Comparison with Edit-based codes}
The edit distance is a more stringent metric compared to Hamming distance. As such, with the equal minimum distance limit, codes constructed with edit distance are likely to have larger energy gaps due to the fewer components than Hamming-based codes. Therefore, for a fair comparison, we make the constructed SS-based codes have the same code sizes with the edit-based codes using expurgation. Table~\ref{Tab:distance effect 2} shows that for all lengths considered, SS-based expurgated codes have larger free energy gaps than edit-based codes under the assumption of the same code sizes, which implies that the SS-based codes have better cross-hybridization performance.




 


\subsubsection{Comparison with codes in \cite{zhang2013novel}}
We set the maximum SS constraint $T_{th}$ based on the minimum distance $d$ (out of $n$) that was used in \cite{zhang2013novel}. We generate each DNA code with maximum SS $T_{th}=1-\frac{d}{n}$ with values of parameter $d$ and sequence length $n$ same as \cite{zhang2013novel}. All codes generated follow $\alpha_1=\alpha_2=1;\beta_1=\beta_2=0$, where a biased similarity weight is allocated to the consecutively identical G/C base-pairs. The free energy gap $\delta$ and the size of the codes are shown in Table~\ref{Tab:comp1} and Table~\ref{Tab:comp2}, respectively. The parenthetical data are from \cite{zhang2013novel}.

\begin{table}
\setlength{\tabcolsep}{1.4mm}
\centering 
\caption{Comparing the free energy gap of DNA codes} \label{Tab:comp1}
\begin{tabular}{|l|l|l|l|l|l|}
\hline
\diagbox{$n$}{$d$} & 4 & 5 & 6 & 7 & 8\\ \hline
4 & 2.7 (1.54) & - & - & - & -\\ \hline
5 & 3.3 (1.94) & 4.5 (2.45) & - & - & -\\ \hline
6 & 2.4 (1.73) & 5.5 (3.04) & 4.5 (3.47) & - & -\\ \hline
7 & 2.1 (1.94) & 4.1 (2.69) & 6.3 (3.89) & 6.5 (4.36) & -\\ \hline
8 & 2.3 (2.59) & 4.1 (3.25) & 6.0 (3.98) & 7.5 (5.12) & 7.5 (5.59)\\ \hline
\end{tabular}
\end{table}


Table~\ref{Tab:comp1} shows that all generated codes except $\mathcal{C}(n=8,d=4)$ have better cross-hybridization performance (i.e., the free energy gap $\delta$) over the codes generated in \cite{zhang2013novel}. However, with a 11\% decrease of the energy gap (i.e., 2.3 versus 2.59), the size of $\mathcal{C}(n=8,d=4)$ increases over 255\% (3.5 fold) (i.e., 32 versus 9 in Table~\ref{Tab:comp2}). Generally, the gap $\delta$ is expected lower with a larger code size $M$. However, Table~\ref{Tab:comp1}  together with Table~\ref{Tab:comp2} indicates that for specific sequence length $n$, the proposed method could generate DNA codes with larger size $M$ and wider gap $\delta$. More specifically, for $n=7$, we generate codes with $M=11, \delta=2.1$ and $M=3, \delta=4.1$ against $M=5, \delta=1.95$ and $M=2, \delta=3.89$ in \cite{zhang2013novel}, respectively. For $n=8$, we generate codes with $M=7, \delta=4.1$ against $M=4, \delta=3.98$. Note that the thermodynamic property of DNA codes in \cite{zhang2013novel} might be worse in practice as they neglected the self-fold hybridizations in their code design and the free energy gap definition. Moreover, unlike \cite{zhang2013novel}, our codes satisfy the maximum homopolymer run/continuity constraint that is desirable for achieving better stability, which explains the smaller sizes of codes for few cases in Table~\ref{Tab:comp2}.

\begin{table}
\setlength{\tabcolsep}{3.3mm}
\centering 
\caption{The comparison on the size of DNA codes} \label{Tab:comp2}
\begin{tabular}{|l|l|l|l|l|l|}
\hline
\diagbox{$n$}{$d$} & 4 & 5 & 6 & 7 & 8\\ \hline
4 & 1 (1) & - & - & - & -\\ \hline
5 & 1 (2) & 1 (1) & - & - & -\\ \hline
6 & 5 (5) & 1 (2) & 1 (1) & - & -\\ \hline
7 & 11 (5) & 3 (4) & 1 (2) & 1 (1) & -\\ \hline
8 & 32 (9) & 7 (6) & 2 (4) & 1 (2) & 1 (1)\\ \hline
\end{tabular}
\end{table}

\section{Conclusion}
We have introduced a new model for designing DNA codes with controlled cross-hybridization performance. This SS model weighs the significance of the undesirable similarity against the desirable similarity, leveraging the fact that bias exists in the hybridization affinities between individual oligo and its RC pair. An improved BAC and a biased similarity weight allocation were incorporated for the realization of SS. Based on the proposed model and a sorting-based search algorithm, DNA codes with different sequence lengths and maximum SS were generated, while satisfying several combinatorial constraints. The free energy gaps and code sizes of these codes imply that the proposed SS presents better approximation towards the thermodynamic property than the traditional models.







\bibliographystyle{IEEEtran}

\bibliography{references}


\end{document}